\documentclass[conference]{IEEEtran}
\IEEEoverridecommandlockouts

\usepackage{cite}
\usepackage{amsmath,amssymb,amsfonts}
\usepackage{algorithmic}
\usepackage{graphicx}
\usepackage{textcomp}
\usepackage{xcolor}
\def\BibTeX{{\rm B\kern-.05em{\sc i\kern-.025em b}\kern-.08em
    T\kern-.1667em\lower.7ex\hbox{E}\kern-.125emX}}

\usepackage{xspace}
\usepackage{fontawesome}
\usepackage{multirow}
\usepackage{rotating}
\usepackage{tikz}
\usepackage{soul}
\usepackage{makecell}
\usepackage{url}
\usepackage{enumitem}
\usepackage{booktabs}
\usepackage{fontawesome}
\usepackage{pifont}
\usepackage{amssymb}
\usepackage[colorlinks=true, allcolors=black]{hyperref}
\usepackage{tcolorbox}
\usepackage[framemethod=TikZ]{mdframed}
\usepackage{arydshln}

\newcommand{\ie}{\emph{i.e.,}\xspace}
\newcommand{\eg}{\emph{e.g.,}\xspace}

\newcommand{\figref}[1]{Fig.~\ref{#1}\xspace}

\newcommand{\tabref}[1]{Table~\ref{#1}\xspace}

\begin{document}

\title{Code Review is a Conversation:\\Toward Conversational AI Review Assistants}

\author{
Rosalia Tufano \\ \textit{SEART @ Software Institute} \\ \textit{Universit\`{a} della Svizzera italiana (USI)} \\Switzerland
}

\maketitle

\begin{abstract}
AI-based code review tools increasingly promise to help developers inspect pull requests, identify defects, and improve code quality. Yet most current approaches frame code review as a one-shot commenting task: given a diff, the system produces warnings or suggestions. This framing overlooks a central property of modern code review: review is a conversation. Human reviewers do not merely comment on code; they ask questions, explain expectations, negotiate design trade-offs, request evidence, transfer project knowledge, document rationale, and collectively decide whether a change is good enough to integrate. In this vision paper, we argue for conversational AI review assistants: systems that participate in code review as interactive partners rather than static comment generators. Such assistants should identify when conversation is needed, ask grounded questions, respond to developer explanations, summarize unresolved issues, help capture rationale, and know when to abstain or escalate to human reviewers. Such a paradigm shift requires novel evaluation methodologies as well. We outline a research agenda for studying review conversations, designing conversational AI review capabilities, and evaluating their impact on software evolution and maintenance. Our vision reframes AI code review from automated commenting to human-AI sensemaking before integration.
\end{abstract}


\section{Introduction}
Code review is increasingly becoming a target for AI-based automation. Recent advances in AI have enabled tools that can inspect diffs, generate review comments, identify potential defects, suggest improvements, and even propose patches \cite{li:fse2022,patanamon:icse2022,tufano:icse2022,li.l:esecfse2022,hong:esecfse2022,frommgen:icse-seip2024,lin:tosem2024,chen:fse-c2025}. These tools have the potential to reduce the cost of code review, and recent work in the literature documented the adoption of Large Language Models (LLMs) as co-reviewers in open source projects \cite{Tufano:msr2024}.

Most current AI-based code review approaches implicitly frame code review as a one-shot commenting task. Given a code diff, the system produces natural language comments (as a human reviewer would do) reporting quality issues and/or explicitly recommending code changes \cite{li:fse2022,tufano:icse2022,li.l:esecfse2022,hong:esecfse2022,lin:tosem2024,chen:fse-c2025}. While such a framing is certainly useful, it assumes that the primary value of review lies in producing comments about code. However, in practice, human reviewers do much more than comment. They ask whether a behavior is intended, request tests to support a claim, explain project conventions, negotiate design alternatives, discuss trade-offs, transfer architectural knowledge, and document the rationale behind integration decisions \cite{bacchelli:icse2013}. Code review is therefore not merely a sequence of comments attached to lines of code, it is a collaborative conversation around a change.

A comment-only framing can lead to AI review assistants that are technically impressive but poorly aligned with the actual developers' workflow: A tool that generates many plausible comments may still fail to help developers reach better decisions. This may also partially explain recent evidence in the literature suggesting that LLM-supported code review can be counterproductive both in terms of review quality (\ie ability to spot quality issues) and cost (\ie time spent reviewing) \cite{tufano:icse2025}.

Conversely, a useful AI review assistant may sometimes produce no direct warning at all. Its most valuable contribution may be to ask a well-grounded question pushing the developer to better reflect on their change, identify an unresolved assumption, or suggest what evidence would make a change easier to trust.

In this vision paper, we argue that the next generation of AI code review tools should move beyond automated commenting toward conversational AI review assistants. By conversational, we do not simply mean chat-based interfaces. Rather, we refer to systems that can participate in the communicative work of code review: identifying when a change requires discussion, asking context-aware questions, responding to developer explanations, helping reviewers reason about tradeoffs, capturing rationale, and knowing when to abstain or escalate to humans. Such a vision is supported by the following statement, vastly supported by evidence in the literature (see \eg \cite{bacchelli:icse2013}):

\begin{quote}
\begin{center}
	\emph{The long-term value of code review is not only the final accepted patch, but also the knowledge produced during review.}
\end{center}
\end{quote}

Realizing this vision requires a shift in how the research community studies, designs, and evaluates AI support for code review. If AI review is treated as comment generation, evaluation naturally focuses on whether generated comments resemble human comments, detect known defects, or receive positive ratings. This is what still happens nowadays even with the newest benchmark aimed at assessing agenting-based code review \cite{lu2025deepcrceval,pereira:2026,zhang2026code,zhong2026human,naik:2025crscore}.

Shifting to conversational review assistance requires broader evaluation questions:
Does the assistant help developers clarify intent? Does it reduce uncertainty? Does it lead to better tests, better explanations, or better design decisions? Does it know when not to intervene? These questions cannot be answered by static benchmarks alone.


This paper outlines a research agenda for conversational AI review assistants. We discuss the limitations of current one-shot AI review paradigms, both in terms of technique design and their evaluation. Then, we define the capabilities needed for conversational review assistance. Finally, we identify key research challenges in modeling review conversations, designing useful AI interventions, evaluating multi-turn human-AI review interactions, and integrating such assistants into real development workflows.

\section{Limitations of One-Shot AI Code Review}
\label{sec:limitations}

The dominant framing of AI-supported code review is a one-shot mapping from a code change to review feedback. This framing is not only a modeling choice; it is also reinforced by how datasets are constructed and by how the approaches themself are evaluated. Existing work has made this task concrete and measurable, but the resulting abstraction captures only a narrow slice of what happens in real review conversations. In the following we describe the main limitations in terms of techniques design and evaluation.

\subsection{Technique Design}
\label{sec:limitations:technique}

A key design limitation comes from the way review data is turned into training and benchmarking instances. Many comment-generation approaches mine historical reviews and summarize them as pairs or triplets such as \emph{code change} $\rightarrow$ \emph{review comment}, or \emph{submitted code} $\rightarrow$ \emph{review comment} $\rightarrow$ \emph{revised code} \cite{tufano:icse2022,li:fse2022,li.l:esecfse2022,hong:esecfse2022,lin:tosem2024}. To make the task learnable, datasets often keep comments that can be localized to a method, line, or diff hunk, and in several cases prioritize comments that are followed by an identifiable code change. For example, datasets derived from review-comment resolution or refinement tend to filter out comments that do not lead to a clean local edit, comments involved in multi-comment discussions, or comments whose effect cannot be uniquely mapped to a subsequent revision \cite{tufano:icse2022,li:fse2022,li.l:esecfse2022}. These choices are reasonable for supervised learning, but they also remove many of the interactions that make review conversational.

\begin{figure}[t]
\centering
    \includegraphics[width=0.95\linewidth]{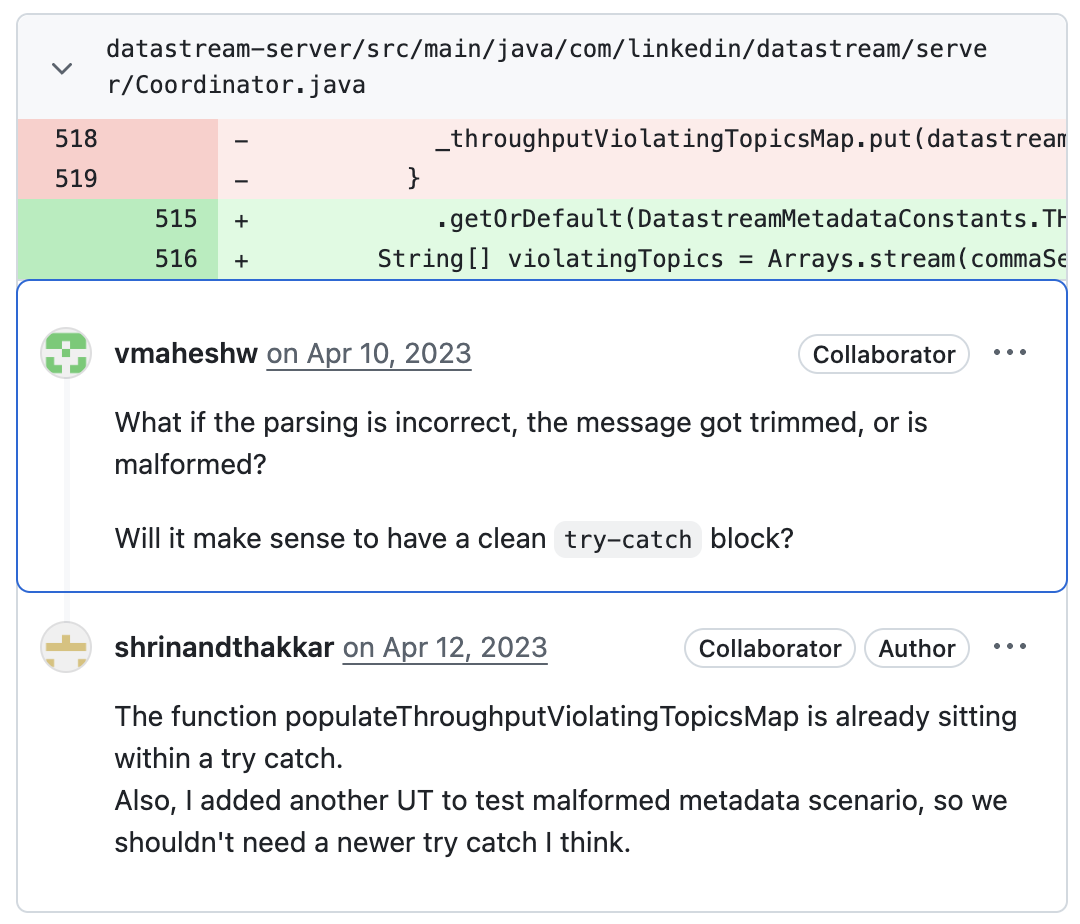}
    \vspace{-8pt}
    \caption{Example of clarification comment asked by the reviewer that does not require any code change: \url{https://github.com/linkedin/brooklin/pull/930}}
    \label{fig:example1}
    \vspace{-30pt}
\end{figure}

The filtered-out cases are precisely those that matter for the vision of conversational review assistance. A reviewer may ask for clarification and receive an explanation without any code change. \figref{fig:example1} shows a concrete example of such a scenario taken from the \texttt{linkedin/brooklin} open source project.

A discussion may end with a design trade-off being accepted, an assumption being documented, or a maintainer deciding that a concern should be deferred. Other comments may request tests, expose uncertainty, transfer project knowledge, or negotiate scope. If such interactions are excluded because they are not followed by a code edit, models learn from a residue of review in which the conversational work has been stripped away.
This also affects the kinds of behavior current AI reviewers are designed to exhibit. A system trained to reproduce historical comments is encouraged to output a comment whenever the task expects one. It has little representation of alternative review actions: asking a question, waiting for an author explanation, summarizing unresolved points, requesting evidence, escalating to a maintainer, or abstaining. 

A related limitation is that the retained comments are often the most localizable part of the review. In platforms such as GitHub, review discussions include both inline comments attached to specific lines of code and broader pull-request conversation threads. The latter may contain explanations of intent, discussion of scope, coordination among maintainers, decisions to defer concerns, or summaries of why a proposed alternative was rejected. These interactions are often not attached to a precise line of code, and they may not lead to an immediate edit. Yet they are central to how reviewers and authors converge to a decision. Thus, even when datasets preserve review comments, they may still capture only the portion of review that is easiest to align with code, while missing the broader conversation in which much of the sensemaking happens.

Recent agent-based systems broaden the implementation of AI review, but not necessarily the interaction model. A multi-agent pipeline may decompose the task into detection, localization, and repair steps \cite{chen:fse-c2025}, and recent benchmarks may evaluate agents in more realistic pull-request settings \cite{pereira:2026,zhang2026code,zhong2026human}. Yet the external behavior is still usually centered on producing review feedback to be judged as useful, correct, or defect-relevant. Internal multi-step reasoning is not the same as participating in a developer-reviewer conversation. Thus, the design limitation is not simply that current systems need stronger models. It is that the field has largely operationalized the reviewer role as producing comments from filtered review artifacts. Moving toward conversational AI review assistance requires datasets and systems whose unit of analysis is the review interaction: the initial change, the comments, the replies, the revisions, the unresolved issues, the decisions, and the rationale that survives after merge.

\subsection{Evaluation}
\label{sec:limitations:eval}

The same abstraction shapes evaluation. When the task is comment generation, the generated comments become the units of assessment. Early and recent systems have therefore been evaluated by running them on past pull requests and comparing the generated comments against historical human comments. This has been done via exact match (\ie the generated comment is identical to the one written by the human reviewer) and textual similarity metrics such as BLEU \cite{papineni:acl2002} and ROUGE \cite{lin:tsbo2004}, \cite{tufano:icse2022,li:fse2022,li.l:esecfse2022,hong:esecfse2022,lin:tosem2024}. These metrics are scalable, but they inherit a questionable assumption: that the comment a human happened to write is the target that an AI reviewer should reproduce. In real review, several different interventions may be equally valid, and the most useful one may not resemble the historical comment at all.

Recent work has started to challenge text-similarity evaluation. DeepCRCEval \cite{lu2025deepcrceval} revisits the evaluation of review comment generation, moving the focus from ``\emph{does the generated comment look like the dataset comment?}'' to ``\emph{would this comment help in finding quality issues?}''. This is done by exploiting LLM evaluators and looking at the comment from several quality aspects (\ie readability, relevance, explanation clarity, problem identification, actionability, completeness, specificity, contextual adequacy, and brevity). CRScore \cite{naik:2025crscore} similarly moves toward reference-free evaluation, by checking how much the generated comments are close to potential issues that can be inferred by automatically analyzing the code diff to be reviewed (\eg the addition of a new loop may generate inefficiencies). These are not necessarily comments that a developer would wrote, but potential issues inferred from the diff using LLMs plus static-analysis tools. AI-generated comments well aligned with those potential issues, are positively judged. This makes the assessment independent from the wording used by human reviewers to report quality issues. 

While DeepCRCEval and CRScore represent important steps ahead in the evaluation of A-based code generation, they still evaluate review feedback as an isolated artifact: whether a generated comment identifies a valid issue in the code.

Agent-oriented benchmarks make a similar move from textual resemblance toward actionable feedback. CR-Bench \cite{pereira:2026} evaluates AI code review agents in terms of preventable defects, issue resolution, trustworthiness, developer acceptability, and factuality. c-CRAB \cite{zhang2026code} constructs review-agent tasks from human review feedback and executable tests, retaining issues that can be objectively checked. These benchmarks are valuable because they make evaluation more concrete and closer to real defects. However, they still assess review primarily as the production of defect-relevant feedback. Conversational phenomena (\ie clarifying intent, negotiating trade-offs, producing evidence, documenting rationale, or deciding that no intervention is needed) remain outside the main evaluation target.


A broader evaluation methodology should measure both conversational value and cost. Useful dimensions include whether the assistant helps participants converge on a decision, surfaces unresolved assumptions, produces evidence, preserves rationale, avoids redundant feedback, and remains calibrated in when and how it intervenes. Without such dimensions, benchmarks will continue to reward systems that generate plausible comments, even when they do not improve code review as a maintenance practice.

\begin{table*}
\caption{Examples of conversational actions for AI review assistants.}
\label{tab:actions}
\centering
\footnotesize
\begin{tabular}{p{0.25\columnwidth}p{0.62\columnwidth}p{0.5\columnwidth}p{0.47\columnwidth}}
\toprule
\textbf{Action} & \textbf{Purpose in review} & \textbf{Example behavior} & \textbf{Failure if absent} \\
\midrule
Ask & Surface uncertainty or hidden assumptions. & Ask whether a behavior is intended for an edge case or legacy path. & Assumptions remain implicit. \\\midrule
Request evidence & Make a claim or decision verifiable. & Suggest a regression test, benchmark, log, or documentation update. & Debate remains opinion-based. \\\midrule
Explain & Transfer project knowledge or make expectations explicit. & Point to a local convention or prior decision relevant to the change. & Rationale is lost. \\\midrule
Compare & Support design trade-off discussion. & Contrast two implementation choices and their maintenance implications. & Alternatives are underspecified. \\\midrule
Summarize & Preserve the state of the discussion. & List what has been resolved and what remains open before merge. & Threads close ambiguously. \\\midrule
Escalate & Recognize human authority or ownership. & Recommend confirmation from a maintainer for an architectural decision. & AI oversteps authority. \\\midrule
Abstain & Avoid noise and unnecessary burden. & Remain silent when a concern is weak, duplicated, or outside confidence. & Over-commenting and fatigue. \\
\bottomrule
\end{tabular}
\end{table*}

\begin{table}
\caption{Examples of review comments illustrating conversational actions.}
\label{tab:examples}
\centering
\begin{tabular}{p{0.14\columnwidth}p{0.7\columnwidth}}
\toprule
\textbf{Action} & \textbf{GitHub example} \\
\midrule
Ask &
In \texttt{NVIDIA/NeMo-speech-data-processor} PR \#63 \cite{nemo_pr63}, a reviewer asks: ``Where do you use this docker compose file? Is it possible to run the scripts without it?''\\
\midrule
Request evidence &
In \texttt{grpc-ecosystem/grpc-gateway} PR \#3646 \cite{grpc_gateway_pr3646}, a reviewer asks: ``The change looks reasonable, but I'm not sure I understand what the testdata changes do. Could you add a test to \texttt{template\_test.go}? Or perhaps add an example of this circumstance\ldots''\\
\midrule
Explain &
In \texttt{nix-community/stylix} PR \#1892 \cite{stylix_pr1892}, a reviewer explains the expected local pattern and points the author to the relevant project documentation.\\
\midrule
Compare &
In \texttt{ethereum/solidity} PR \#12978 \cite{solidity_pr12978}, a reviewer suggests an alternative: ``Alternatively, you could reuse the same field in \texttt{Switch}\ldots'' \\
\midrule
Summarize &
In \texttt{servo/fontsan} PR \#46 \cite{fontsan_pr46}, a reviewer summarizes the discussion with: ``So to summarize the workflow is as follows, right?'' and lists the agreed steps.\\
\bottomrule
\end{tabular}
\end{table}

\section{Conversational Actions for AI Review Assistants}
\label{sec:capabilities}

If code review is a conversation, an AI review assistant should be designed around conversational actions, not only around comment generation. The goal is not to replace human reviewers or to automate the merge decision. Rather, the assistant should help the review process move toward a better-supported integration decision by contributing at moments where conversation can improve understanding, evidence, or coordination.

\tabref{tab:actions} summarizes examples of conversational actions that could become first-class capabilities of future AI review assistants. These actions differ from conventional comment generation because they are not all attempts to point out an issue. Some aim to elicit missing information, some help developers reason about alternatives, some manage the state of the discussion, and some deliberately avoid intervention. This broader action space is essential if AI support is to fit the way human review actually unfolds. \tabref{tab:examples} reports, concrete examples of actions performed by human reviewers for the top-five actions listed in \tabref{tab:actions}; the last two actions are instead more typical of AI agents. Note that all defined actions could be suitable for the definition of trajectories in agentic code review.

The proposed framing also changes the role of abstention and escalation. In a comment-generation task, abstention is often treated as failure to produce output. In conversational review, abstention is a necessary capability: the assistant should remain silent when an intervention would add little value or distract from more important discussion. Similarly, escalation is not a failure of automation but a recognition that some decisions require human ownership and/or domain expertise.

These capabilities expose the design gap left by one-shot AI review. A conversational assistant needs to decide when to intervene, what type of intervention is appropriate, how to adapt to responses, and how to preserve the knowledge produced by the review. These requirements motivate the research agenda that follows.

\section{Operationalizing Conversational Review Assistance}
\label{sec:operationalizing}

A conversational review assistant should not be understood as a chatbot attached to a pull request, but as a stateful participant in the review process. Its input is not only the current diff, but also the evolving review state: existing comments, author replies, pushed revisions, test results, linked issues, ownership information, and project conventions. Its output is not necessarily a defect report, but an intervention selected to improve the state of the review.

We can model this process as a loop with five stages. First, the assistant \emph{observes} the current review state and identifies open uncertainties, missing evidence, unresolved comments, duplicated concerns, or decisions requiring human ownership. Second, it \emph{estimates} whether an intervention is likely to improve the review or merely add noise. Third, it \emph{selects an action} from a broader action space, such as asking a clarification question, requesting evidence, explaining a convention, comparing alternatives, summarizing the discussion, escalating to a maintainer, or abstaining. Fourth, it \emph{updates} its representation of the review state by incorporating author replies, new commits, tests, and reviewer decisions. Finally, near integration time, it helps \emph{close the rationale} by preserving what was decided, what evidence supported the decision, and which concerns were intentionally deferred or rejected.

The timeout example above illustrates why this stateful loop matters. A conventional reviewer model stops after producing a plausible warning. A conversational assistant can instead ask whether the timeout applies to retries, observe the author's clarification, inspect whether the new implementation is consistent with that clarification, request a regression test if evidence is missing, and later summarize that the intended behavior has been documented. The useful artifact is therefore not only a comment, but a better-supported integration decision.

This operationalization also makes evaluation more concrete. An assistant asking for a regression test should be assessed not only by whether the request sounds plausible, but by whether the request leads to useful evidence, whether the author understands the concern, and whether the resulting thread documents the intended behavior. Similarly, an assistant that abstains should be credited when the available evidence is weak, the concern duplicates an existing comment, or the decision clearly belongs to a human maintainer. Table~\ref{tab:evaluation} summarizes candidate evaluation dimensions and possible operationalizations that follow from this interaction-centered view.

These measures are imperfect proxies rather than a fixed benchmark specification. Their purpose is to make explicit that conversational review assistance should be evaluated by its effect on the review state and outcome, not only by the plausibility of its first generated comment. We also do not claim that conversational AI interventions are always beneficial: poorly timed, overly verbose, or socially inappropriate interventions may increase review burden. For this reason, abstention, calibration, and cost-sensitive evaluation must be first-class parts of the research agenda.

\section{Research Agenda}
\label{sec:agenda}

Conversational AI review assistance requires progress on four connected research problems.

\subsection{Modeling Review Conversations}
\label{sec:agenda:model}

The first challenge is to build data and models at the level of review threads rather than isolated comments. A thread-level representation should include comments, replies, code revisions, author explanations, linked issues, tests added during review, decisions to defer or reject concerns, and the final integration outcome. This would make it possible to study not only what reviewers say, but also how discussions evolve and how they are resolved. Open questions include how to identify conversational turns, how to represent parallel discussions by multiple reviewers, and how to label outcomes such as ``resolved by test'', ``resolved by explanation'', ``accepted trade-off'', ``deferred to follow-up'', or ``unresolved at merge''. A concrete research direction is to develop taxonomies for review-state transitions, that could also help in shaping trajectories for agentic code review: what happened after a comment, why the thread was considered resolved, and which knowledge should remain visible to future maintainers.

\subsection{Designing Useful AI Interventions}
\label{sec:agenda:design}

The second challenge is to decide when and how an assistant should intervene. This is an action-selection problem, not only a generation problem. At any point in a review thread, the assistant may ask a question, request evidence, summarize the state of the discussion, suggest a next step, escalate to a human reviewer, or abstain. Designing such policies requires modeling uncertainty, potential benefit, and intervention cost. It also requires avoiding two opposite failures: over-commenting, which adds noise and burden, and under-intervening, which leaves useful support unused. Important questions include when a question is preferable to a recommendation, how to estimate whether an intervention will reduce uncertainty, and how to distinguish weak concerns from concerns that should be escalated to project experts.

\begin{table}[h]
\caption{Candidate measures for evaluating conversational review assistance.}
\label{tab:evaluation}
\centering
\footnotesize
\begin{tabular}{p{0.31\columnwidth}p{0.59\columnwidth}}
\toprule
\textbf{Dimension} & \textbf{Possible operationalization} \\
\midrule
Uncertainty reduction & Unresolved assumptions before and after intervention; author-confirmed clarification. \\\midrule
Evidence production & Tests, benchmarks, logs, documentation, or issue links added during review. \\\midrule
Rationale capture & Final thread or merge summary explains the accepted decision and rejected alternatives. \\\midrule
Developer burden & Extra comments, response effort, perceived interruption, or review time. \\\midrule
Calibration & Fraction of interventions judged useful versus unnecessary; appropriate abstentions. \\\midrule
Conversation progress & Whether the thread reaches resolution with fewer open concerns or redundant exchanges. \\\midrule
Maintenance outcome & Post-merge rework, revert rate, or later confusion around the reviewed code. \\
\bottomrule
\end{tabular}
\end{table}

\subsection{Evaluating Human-AI Review Conversations}
\label{sec:agenda:eval}
A conversational AI review assistant should not be evaluated only by whether its comments match human comments or detect known defects. It should be evaluated by whether the interaction leads to better review outcomes. \tabref{tab:evaluation} lists relevant dimensions that aim at capturing conversation progress, reduced uncertainty, evidence produced, rationale captured, developer burden, trust calibration, and post-merge rework. These dimensions require methods beyond static benchmarks: controlled studies with developers, replay studies over historical review threads, and longitudinal field studies of teams adopting conversational assistants. 

\subsection{Integration into Real Development Workflows}
\label{sec:agenda:integration}

The fourth challenge is integration. Conversational assistants must fit into existing review interfaces and social norms. They must decide whether to post publicly in a PR, assist a reviewer privately, or summarize discussion at specific decision points. They must coexist with human reviewers and with the spontaneous use of LLMs as co-reviewers already observed in open-source projects \cite{Tufano:msr2024}. They also raise practical questions about privacy, ownership of review data, and responsibility for integration decisions when AI has actively participated in the thread. A practical design question is therefore not only what the assistant should say, but where, to whom, and with what level of authority it should say it. 

\section{Conclusion}
\label{sec:conclusion}

AI support for code review has largely been built around the one-shot production of review comments. This framing has enabled progress, but it also filters away much of what makes review valuable for software maintenance: questions, explanations, negotiation, evidence production, and rationale capture. As a result, current systems and benchmarks risk optimizing for plausible feedback rather than for better review conversations.

In this vision paper, we have argued for a shift toward conversational AI review assistants: systems that help developers reach clearer, better-supported integration decisions. Such assistants should not be judged by how many comments they generate, but by whether they help teams surface assumptions, resolve uncertainty, produce evidence, document rationale, and leave the codebase easier to understand and maintain. The next step for AI-based code review is therefore not simply better commenting, but better support for the conversations through which software evolves.

\section*{Acknowledgment}
We acknowledge the financial support of the Swiss National
Science Foundation for the PARSED project (SNF Project No.
219294).

\bibliographystyle{IEEEtran}
\bibliography{main}

\end{document}